\documentclass[twoside,english,12pt]{elsart}

\usepackage{amssymb}

\usepackage{graphicx}
\usepackage{babel}
\usepackage{multicol}
\usepackage[centerlast]{caption2}

\newcommand{\registered}{\circledR}

\begin{document}

\begin{frontmatter}
\title{The Effect of Gas Ion Bombardment on the Secondary Electron Yield of TiN, TiCN and TiZrV
Coatings For Suppressing Collective Electron Effects in Storage
Rings}

\author{F. Le Pimpec}
\address{Paul Scherrer Institute \\5232 Villigen
Switzerland}

\author{R.E. Kirby\corauthref{cor1}}
\ead{rek@slac.stanford.edu}
\author{F.K. King and M. Pivi}
\address{Stanford Linear Accelerator Center, \\ 2575 Sand Hill Road, Menlo Park CA-94025, USA}

\corauth[cor1]{Corresponding author}

\begin{abstract}
In many accelerator storage rings running positively charged
beams, multipactoring due to secondary electron emission (SEE) in
the beam pipe will give rise to an electron cloud which can cause
beam blow-up or loss of the circulating beam. A preventative
measure that suppresses electron cloud formation is to ensure that
the vacuum wall has a low secondary emission yield (SEY). The SEY
of thin films of TiN, sputter deposited Non-Evaporable Getters and
a novel TiCN alloy were measured under a variety of conditions,
including the effect of re-contamination from residual gas.
\end{abstract}


\journal{Nuclear Instruments and Methods in Physics Research : A }
\end{frontmatter}

\section{Introduction}

The electron cloud effect (ECE) may cause beam instabilities in
accelerator structures with intense positively charged bunched
beams, and it is expected to be an issue for the positron Damping
Ring (DR) of the International Linear Collider (ILC). Reduction of
the secondary electron yield (SEY) of the beam pipe inner wall is
effective in controlling cloud formation. We have previously
measured the secondary electron emission (SEE) from a number of
technical surfaces and coatings used in ring construction
\cite{lepimpec:Nima2005}, including uncoated aluminium alloys
\cite{lepimpec:jvsta2005}. Here, we present SEY ($\delta$)
measurements, after various treatments including ion bombardment,
on TiCN, TiN and two differently-deposited non-evaporable getter
(NEG) TiZrV films on aluminium substrates. All samples were
produced at Lawrence Berkeley National Laboratory (LBNL).

\section{Experiment description and methodology}

The system used to measure the SEY is described in detail in
\cite{lepimpec:jvsta2005}. Measuring techniques included x-ray
photoelectron spectroscopy (XPS) and residual gas analysis (RGA).
Sample processing facilities were heating and ion bombardment.

The SEY ($\delta$) definition  is determined  from
equation~(\ref{equdefinition}). In practice
equation~(\ref{equSEY}) is used because it contains parameters
directly measured in the retarding target potential experiment.

\begin{equation}
\delta = \frac{Number\ of\ electrons\ leaving\ the\
surface}{Number\ of\ incident\ electrons}
\label{equdefinition}
\end{equation}
\begin{equation}
\delta = 1 -\frac{I_T}{I_P}
\label{equSEY}
\end{equation}

I$_P$ is the primary current (the current leaving the electron gun
and impinging on the surface of the sample) and I$_T$ is the total
current measured on the sample ($I_T~=~I_P~-~I_{SE}$). I$_{SE}$ is
the secondary electron current leaving the target.

The SEY is measured, at normal incidence, by using a gun capable
of delivering an electron beam of 0-3~keV, working at a set
current of 2~nA and having a 0.4~mm$^2$ spot size on the target.
The measurement of the SEY is done while biasing the sample to
-20~V. This retarding field repels most secondaries from adjacent
parts of the system that are excited by the elastically reflected
primary beam. The primary beam current as a function of the
primary beam energy is measured and recorded each time before an
SEY measurement, by biasing the target to +150~V, and with the
same step in energy for the electron beam. A fresh current lookup
table is created with each measurement. The SEY measurement, over
the 0-3~keV range, takes around 5~minutes.

In order to study the effect of ion bombardment on the SEY, we
used a micro-focussing scanning gas ion gun (Leybold IQE 12/38).
The gun has two differentially-pumped beam formation stages that
reduce the sample system pressure compared to that inside the
gun's electron-impact ionization chamber (into which the gas is
directly-injected). Ion energies from 250-5000 eV are possible.
Five nines-pure hydrogen or nitrogen gases were used in this
particular set of experiments. In an accelerator, the ions
produced by beam ionization of residual gases, have a spread in
energy.  In one of the ILC damping ring designs (6~km), the impact
energy of the ions is around 140~eV \cite{lanfawang}. Our ion gun
is not designed to work below 250~eV; therefore, we have set the
energy of the test ions to be 250~eV. The modest increase in ion
energy will raise the nitrogen ion (momentum) and hydrogen ion
(chemical) sputter yields from 0.1 to 0.15, for removing
hydrocarbon contamination \cite{Hopf:2003}. The outermost layers
of the aluminium are composed of hydrocarbons and water on top of
native oxide.  All three materials raise the secondary yield
\cite{Halbritter:84}.  The nitrogen momentum sputter yield is
lower for the native aluminum oxide than for the loosely-bound
hydrocarbons and water; however, metal oxides are removable by the
hydrogen chemical sputtering \cite{Hoyt:Pac95}. In our setup, the
conditioning ions, hydrogen  and  nitrogen, are impacting onto the
sample surface at an angle of 35~$^\circ$ from the sample normal,
with an ion density of  $\sim$10$^{10}$ cm$^{-2}$s$^{-1}$. [N.B.
This rate is 8~nA on 1" (2.54~cm) diameter sample] The expected
species content of the beam, for an electron-impact source using
H$_2$ or N$_2$, is 50\% charged (mostly single-charge diatomic)
and 50\% charge-exchanged energetic neutrals \cite{kirby:1980}.
However, the beams will be referred to as "H$_2^+$" or "N$_2^+$".

\medskip
\begin{table}[tbph]
\centering
\caption{Measurement history of air-exposed thin film samples.}
\begin{tabular}{|c|c|c|c|c|c|}
  \hline
   Film &  Measured & Activated & Vacuum & Ion \\
   &  as received & or baked & Recontamination & conditioning \\
  \hline
  TiCN/Al &  Y & 170$^\circ$C - 2H  & Y & H$_2$ 250 eV  \\
   \hline
  NEG A&  Y & 215$^\circ$C - 1.75H  & Y & N$_2$ 250 eV   \\
   \hline
   NEG B &  Y & 212$^\circ$C - 2H  & Y & -   \\
   \hline
  TiN/Al &  Y & -  & - & N$_2$ 250 eV  \\
   \hline
  \end{tabular}
\label{tabHistoryfilms}
\end{table}

The films, deposited on 6063 aluminium alloy substrates, are
listed in Table.\ref{tabHistoryfilms}, along with their treatment
history. The TiZrV NEG films were produced either from an
arc-melted cathode (A) or from a sintered powder cathode (B). The
two different TiZrV deposition cathodes were used in order to
discover which produced dense adherent films of proper
stoichiometry. Both did and the results were consistent with the
SAES films. The composition in NEG films prepared by CERN and SAES
Getters$^{\registered}$ were studied earlier
\cite{lepimpec:Nima2005}. The composition, in at\%, of the
coatings TiCN, NEG A\&B and TiN is listed in
Table.\ref{tabAtomicconc}.

\begin{table}[tbph]
\centering
\caption{Atomic composition (at\%) of the different coatings}
\begin{tabular}{|c|c|c|c|c|c|}
  \hline
   & Ti & Zr & V & C & N \\
  \hline
  TiCN & 12 & - & - & 55 & 33 \\
  TiZrV - A & 29 & 25 & 46 & - & - \\
  TiZrV - B & 33 & 25 & 42 & - & - \\
  TiN & 51 & -& -  & - & 49  \\
  \hline
\end{tabular}
\label{tabAtomicconc}
\end{table}

\section{Results, TiCN}

\begin{figure}[tbph]
\begin{center}
\includegraphics[width=0.7\textwidth,clip=]{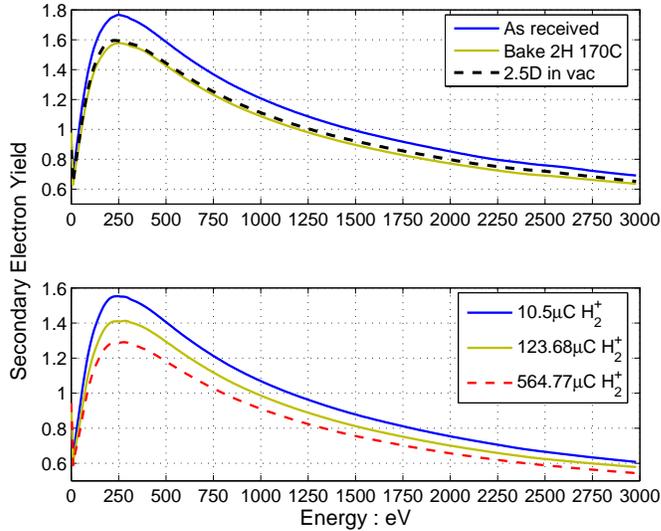}
\end{center}
\caption{SEY obtained on TiCN/Al sample, at normal incidence,
after different accumulated hydrogen ion doses.}
\label{figTiCN}
\end{figure}

This ternary film was chosen to be a possible alternative to TiN
or NEG coatings. It is known that as-deposited titanium nitride
and carbide have a $\delta_{max}$ around or below 1
\cite{crc,Garwin:1987}; however, after deposition and air
exposure, the SEY degrades to such extent that $\delta_{max}$  is
above 1.5 \cite{lepimpec:Nima2005,Garwin:1987,he:EPAC04}. We
wanted  to test whether a ternary alloy would have different
properties when exposed to air than had the pure nitride and
carbide. A film was magnetron sputter-deposited from a TiCN
cathode in Ar/N$_2$ atmosphere onto aluminium sheet. The atomic
film composition, measured by energy-dispersive x-ray
spectrometry, is presented in Table.\ref{tabAtomicconc}. The
results are presented in Fig.\ref{figTiCN}.

The SEY curve and $\delta_{max}$ of TiCN, as-deposited and
air-exposed ("as-received"), and after heating are similar to that
of TiN \cite{Hilleret:HPC02}. Short-term recontamination by
residual gas, at a pressure of ~5.10$^{-10}$~Torr, had a
negligible effect on the SEY.

With respect to ion bombardment behaviour, it is known that a glow
discharge (Argon or Nitrogen) (ArGD or NGD) bombardment on
technical surfaces will sputter-clean the surface to such an
extent that its SEY will be very close from the atomically clean
surface \cite{calder:1986,Padamsee:1979}. We can expect that such
plasma will also work on thin films. However, a GD is not
sustainable in ultra-high vacuum. That implies a high current of
ions impinging the surfaces. It is desirable to reproduce the
impact of hydrogen ions, expected to be created by the circulating
accelerator beam, on the TiCN coating. Hydrogen, of course, was
chosen because it is the highest concentration gas in a baked
vacuum environment, Table.\ref{tabVACpressure}. However, because
the ionization cross section is smaller for H$_2$ than for CO
(mass 28), we also tested the effect of using an analog of this
much-higher mass projectile on the coatings. CO contaminates
vacuum systems with carbon, so equally-massive N$_2$ was
substituted. Chemically-active molecules, like CO, also do
adsorption surface chemistry but we expect that this will be a
negligible effect compared to the sputtering energy available in
the ion beam.

\begin{table}[tbph]
\centering \caption{System partial pressures, before ion
bombardment, at P$_t \sim$2.7 10$^{-10}$~Torr}
\begin{tabular}{|c|c|c|c|}
  \hline
  Mass (amu) & 2 & 16  & 28   \\
  \hline
  Current (10$^{-10}$ A) & 7 & 0.2 &  0.7  \\
  \hline
\end{tabular}
\label{tabVACpressure}
\end{table}

During H$_2^+$ bombardment, the system pressure rose to
1.2~10$^{-8}$ Torr equivalent N$_2$, with  98\% of the spectrum
dominated by H$_2$. Table.\ref{tabH2pressureTiCN} shows an example
of the current recorded by the system RGA during the operation of
the ion gun with hydrogen. Taking into account the sensitivity of
the ion gauge, the true hydrogen pressure is 2.4~10$^{-8}$ Torr
equivalent H$_2$. The effect of the bombardment is shown in
Fig.\ref{figTiCN}. Even at this low ion energy and dose, the
effect of ion bombardment on the SEY is evident.

\begin{table}[tbph]
\centering \caption{System partial pressures, during H$_2^+$
exposure, P$_t \sim$3.6 10$^{-9}$~Torr}
\begin{tabular}{|c|c|c|c|c|}
  \hline
  Mass (amu) & 2 & 16 & 18 & 28   \\
  \hline
  Current (10$^{-9}$ A) & 13.1 & 0.4 & 0.1 & 1.42  \\
  \hline
\end{tabular}
\label{tabH2pressureTiCN}
\end{table}

\begin{figure}[tbph]
\begin{center}
\includegraphics[width=0.6\textwidth,clip=]{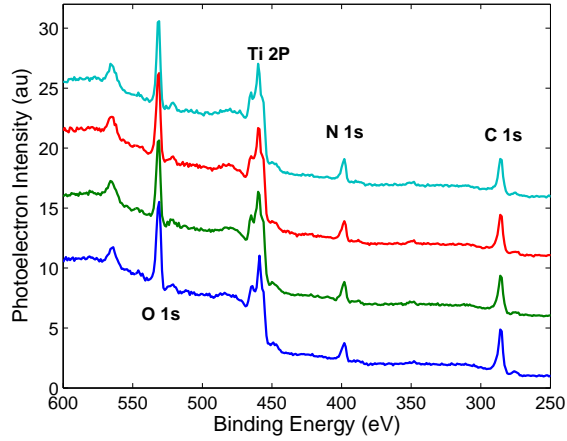}
\end{center}
\caption{XPS survey spectra  of TiCN  following  different
processes. Bottom to top : as received, baked 170$^\circ$ for 2
hours, 2.5 days in vacuum, after 565$\mu$C of H$_2^+$ exposure.
Spectra are vertically displaced for clarity.}
\label{figXPSTiCN}
\end{figure}

XPS-determined  surface  chemistry  showed essentially no changes
as a result of ion bombardment (Figure.\ref{figXPSTiCN}). The SEY is
significantly more sensitive to surface modification (average
escape depth of "true" secondaries $\sim$ 1~nm) than XPS (average
escape depth 3-5~nm).

\medskip

\section{Results, TiZrV}

\begin{figure}[htbp]
\begin{minipage}[t]{.5\linewidth}
\centering
\includegraphics[width=0.95\textwidth,clip=]{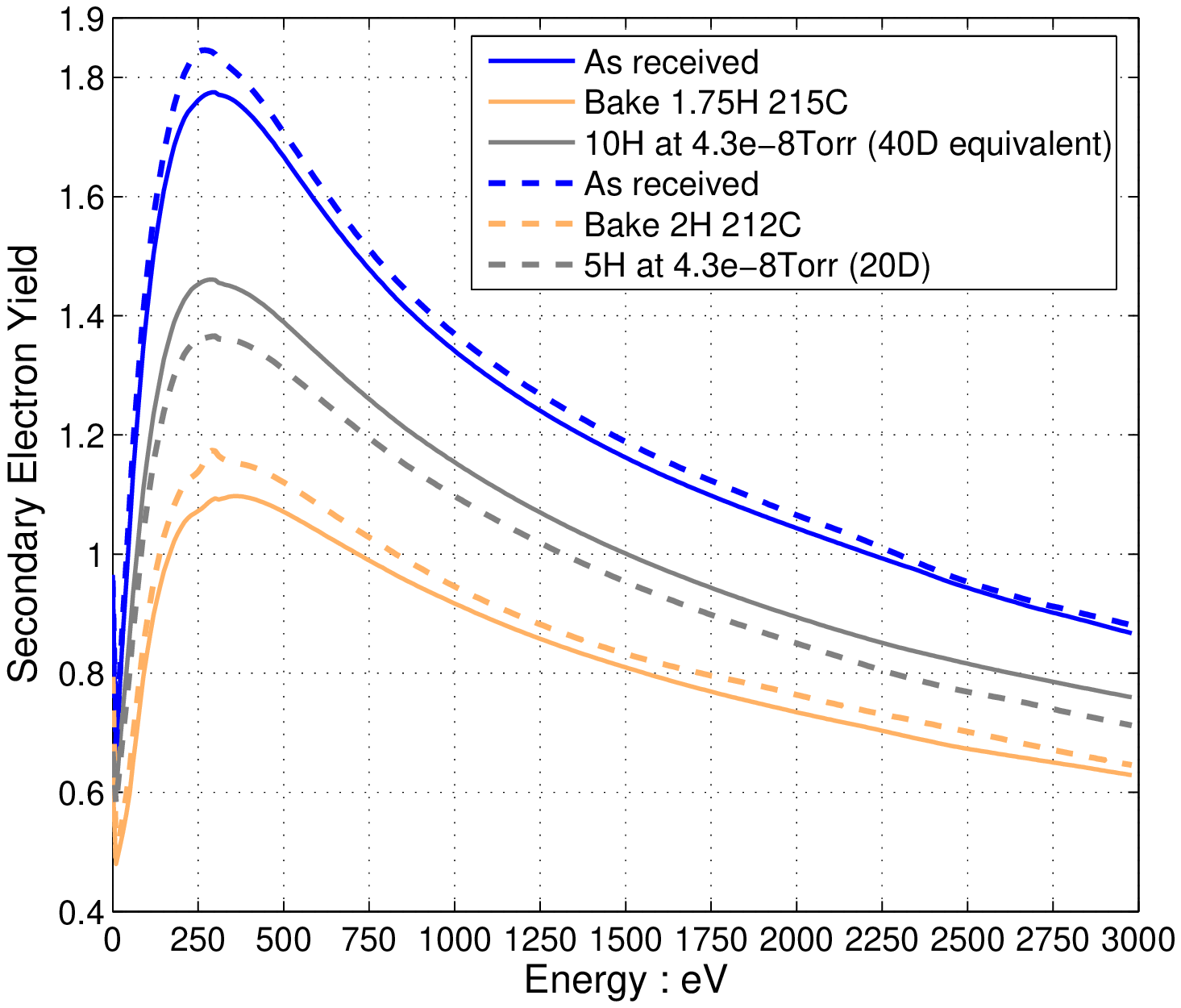}
\end{minipage}%
\begin{minipage}[t]{.5\linewidth}
\centering
\includegraphics[width=0.95\textwidth,clip=]{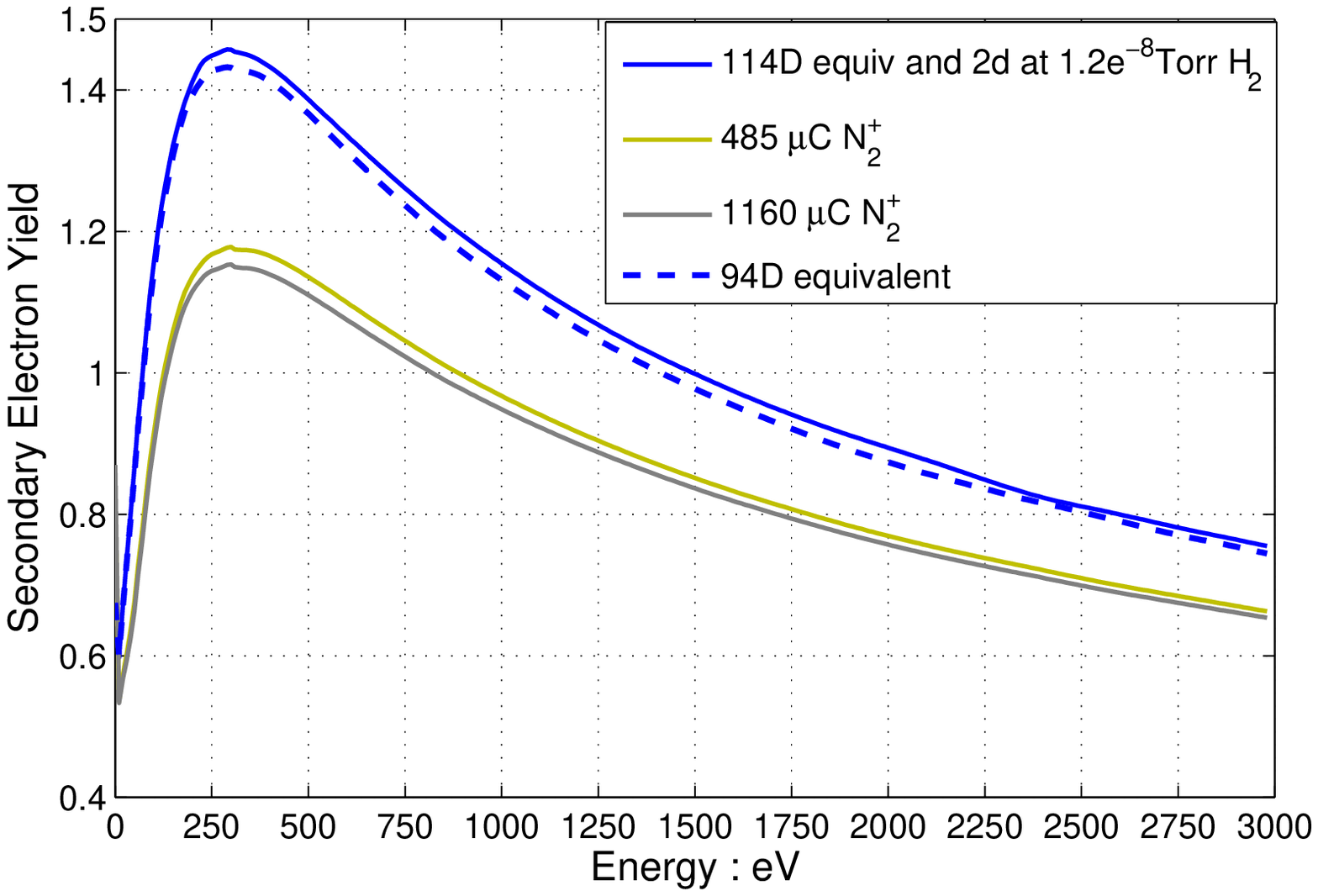}
\end{minipage}
\caption{SEY obtained, at normal incidence, on NEG A (Solid line)
and NEG B (Dashed line), on both right and left figures. Plots are
following the chronology of Table.\ref{tabHistoryfilms}. }
\label{figTiZrVSEY}
\end{figure}

The $\delta_{max}$ values measured for the "as received" NEG
prepared at LBNL, Fig.\ref{figTiZrVSEY}, are slightly below the
values obtained from the earlier-measured CERN and SAES samples
\cite{lepimpec:Nima2005}, 1.8 versus 1.9 and 2.1, respectively.
The SEY values obtained after a 2 hours thermal activation are
similar and slightly below 1.2 \cite{lepimpec:Nima2005}.

Residual gas re-contamination of the surface was also studied. The
sample was transferred to the higher background pressure of the
adjacent load lock chamber.  The load lock was unbaked and the
total pressure, with the sample, was between 2.10$^{-8}$~Torr and
4.8~10$^{-8}$~Torr. This pressure was roughly 100 times above the
average pressure in the measurement chamber. The unbaked residual
gas composition in the load lock chamber
(Table.\ref{tabN2pressureNEG}) is quite different from that of the
measurement chamber, which is dominated by H$_2$
(Table.\ref{tabVACpressure}).  In reference
\cite{lepimpec:Nima2005}, residual gas exposure was done in the
measurement chamber. In this work, it was done in the higher
residual pressure of the load lock chamber, to speed up the
process. Gas exposure, in this work, is listed in "day
equivalents" of the measurement chamber for easy comparison with
the earlier work \cite{lepimpec:Nima2005}. Scaling linearly, 1~s
spent in the load lock correspond to 100~s spent in the
measurement system, for a total pressure in the load lock being
100 times the one in the measurement system. As the saturation is
due to sticking of oxygen based molecules, in UHV, and as their
sticking factor, on a freshly activated NEG, is at least 10 times
the one of hydrogen, comparing directly the total pressure despite
being not rigorous is reasonable.

\begin{table}[tbph]
\centering
\caption{Load lock partial pressures, during
re-contamination , P$_t$ $\sim$4.6~10$^{-8}$~Torr}
\begin{tabular}{|c|c|c|c|c|c|c|c|}
  \hline
  Mass (amu) & 2 & 16 & 17 & 18 & 28 & 32 & 44  \\
  \hline
  Current (10$^{-12}$ A) & 1.2 & 0.3 & 3 & 10 & 2.9 & 0.2 & 1.5  \\
  \hline
\end{tabular}
\label{tabN2pressureNEG}
\end{table}

The SEY  max  of  both  NEG samples  seems to saturate around 1.42
after 40~days equivalent exposure, dashed line in the left upper
plot of Fig.\ref{figTiZrVSEY}. For NEG~A, there is no difference
between the SEY obtained at 40 days equivalent and the one
obtained at 114~days equivalent (not shown) as well as the SEY
obtained for the NEG being in the load lock for an 114~days
equivalent and 2~days inside the measurement chamber during
H$_2^+$ exposure of the TiCN, lower left plot in
Fig.\ref{figTiZrVSEY}. NEG~B show a slight evolution between the
20~days equivalent and 94~days equivalent. However, the SEY max
measured is almost equal to the one obtained for NEG~A.

With respect to the type of gas molecule that causes the rise in
SEY, to saturation (1.42, in our case), Scheuerlein
\cite{Scheuerlein:2002} found that CO or water gave the same
saturation value (1.4, in his case), and that it is the
oxygen-active molecules that matter. The gas adsorption growth
rate on the surface is determined by the partial pressure of the
gas and by the surface coverage-dependant sticking coefficient of
each gas species present. Water vapor is the dominant gas in our
load lock and CO in the UHV chamber. Both oxidized the surface to
similar condition.

N$_2^+$ ion-bombardment was performed on NEG~A. The $\delta_{max}$
decreases and levels  off  just  below  1.2,
Fig.\ref{figTiZrVSEY}. H$_2^+$ ion-bombardment was not done on
either NEG sample because hydrogen diffusion and embrittlement
effects could change the sample morphology \cite{Chiggiato,Yulin}.
Therefore, the NEG was only N$_2^+$ ion-bombarded.

\begin{figure}[htbp]
\begin{minipage}[t]{.5\linewidth}
\centering
\includegraphics[width=0.95\textwidth,clip=]{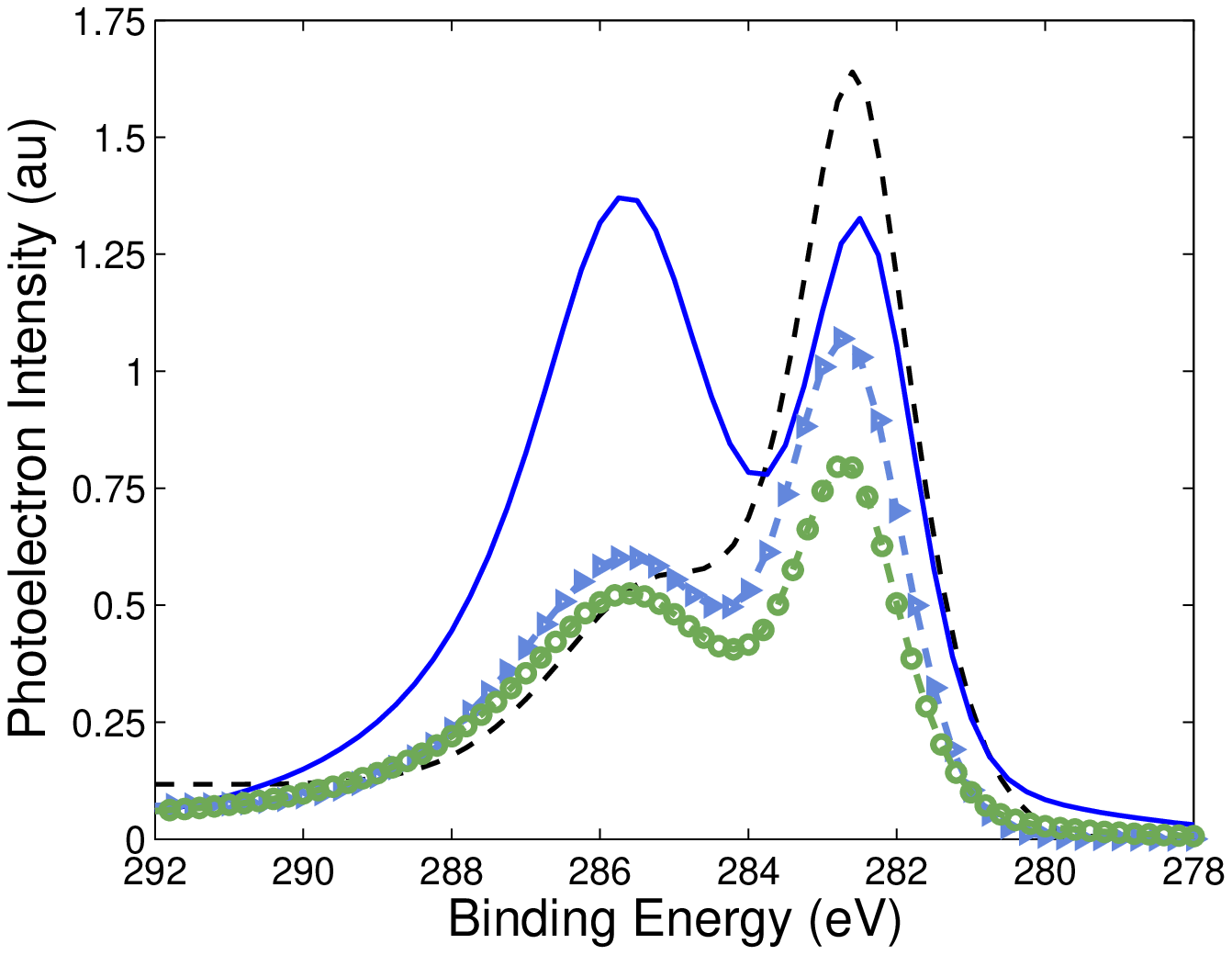}
\end{minipage}%
\begin{minipage}[t]{.5\linewidth}
\centering
\includegraphics[width=0.95\textwidth,clip=]{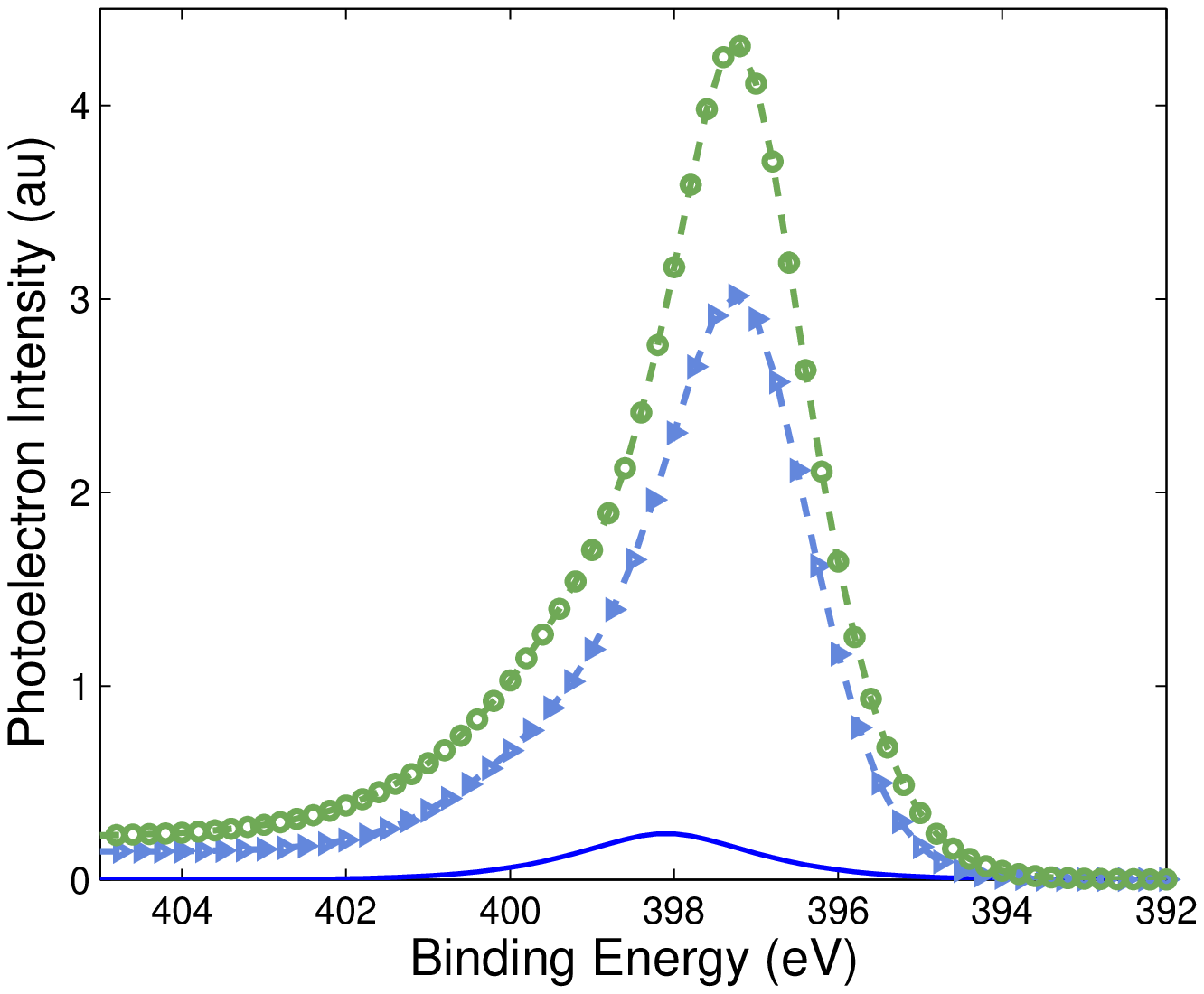}
\end{minipage}
\caption{XPS of C1s (left plots) and N1s peaks (right plots) of
TiZrV/Al, NEG A, after activation (dashed line), after spending
114~days equivalent in the load lock and after 2~days in an
H$_2$-dominated environment (solid line), exposed to 485~$\mu$C of
N$_2^+$ ion bombardment (triangles) and exposed to 1160~$\mu$C of
N$_2^+$ ion bombardment (open circles)}
\label{figXPSNEGACN}
\end{figure}

The NEG A surface chemistry was measured with XPS following the
processes listed in Fig.\ref{figTiZrVSEY}. XPS analysis from the
as-received state (air-exposed getter film) to the activated state
has already be discussed elsewhere
\cite{lepimpec:Nima2005,Scheuerlein:2002,lozano:2000}. XPS results
on re-contamination in  vacuum  has  also  been documented  with
regard  to  the variation  of  the  SEY \cite{lepimpec:Nima2005}.
We focus here, instead, on the evolution  of  the  surface under
N$_2^+$ ion bombardment, Fig.\ref{figXPSNEGACN}.

After activation, the TiZrV  displays C1s carbon peaks at 283~eV
and 285~eV binding energy (BE). They are the marks of carbide and
of amorphous carbon.  As contamination by the residual gas occurs,
mainly  by dissociative adsorption of oxygen on the metallic NEG,
the 285~eV peak increases and get broadened to higher BE. The
broadening is due to contribution of a carbon oxide peak, which
sometimes can be well separated from the 285~eV peak. Upon N$_2^+$
ion-bombardment, surface molecules/atoms are sputtered away and
the carbon and oxygen peak intensities reduce (O 1s data not shown
here). Atoms may also be lost from the surface by diffusion, for
example, into hot NEG. However, in the latter case, the surface
chemistry  is  not changed significantly, as evidenced by the lack
of chemical shifts in the peaks' BE. In other words, the 250~eV
N$_2^+$ ions do not simply induce bonds re-arrangements. Finally,
Fig.\ref{figXPSNEGACN}, right plot, shows that nitrogen gets
implanted in the NEG and adsorbed. The rather broad spectrum of
chemical valence states shows that the nitrogen is populating at
least the outer five nm of surface.

\begin{figure}[tbph]
\begin{center}
\includegraphics[width=0.6\textwidth,clip=]{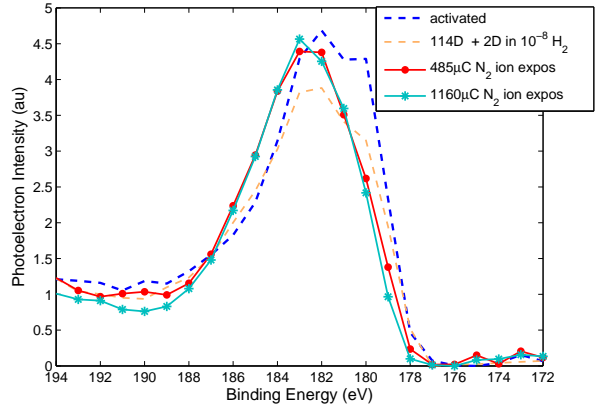}
\end{center}
\caption{Evolution of the chemistry of Zr 3d}
\label{figXPSNEGAZr}
\end{figure}

The XPS Ti peaks (2p$_{1/2}$ and 2p$_{3/2}$) do not show changes.
However, there was a slight evolution in the Zr3d,
Fig.\ref{figXPSNEGAZr}.  Zr has two peaks, located at 178.8 eV and
181.1 eV BE, 3d$_{5/2}$ and 3d$_{3/2}$ respectively.  During
nitrogen ion bombardment, the intensity of the 3d$_{5/2}$ peak
diminishes and the spectrum shifts to slightly higher BE, perhaps
indicative of morphology changes.

\section{Results, TiN}

\begin{figure}[tbph]
\begin{center}
\includegraphics[width=0.6\textwidth,clip=]{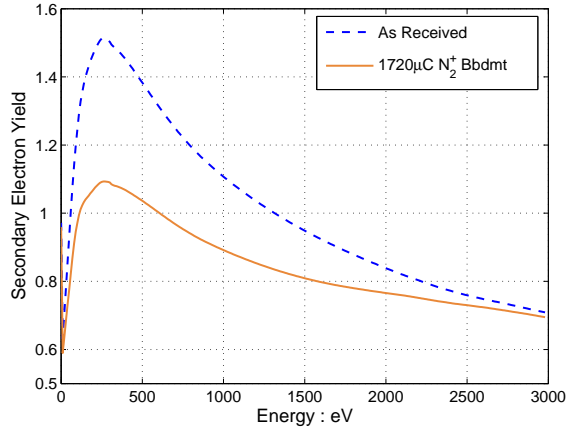}
\end{center}
\caption{SEY at normal incidence, of coated TiN/Al, as received
(dashed line) and after 1720~$\mu$C N$_2^+$ ion bombardment
(solid)}
\label{figTiNhalfflat}
\end{figure}

TiN coating is commonly used to mitigate multipacting in
accelerator and storage ring structures \cite{Welch:1472}. Its SEY
properties are rather well known, but not exhaustively so, for a
wide range of substrate material, deposition thicknesses and
processing conditions
\cite{lepimpec:Nima2005,Garwin:1987,Hilleret:HPC02,kirby:2001,galan:esa03}.
This sample was N$_2^+$ ion-bombarded (Fig.\ref{figTiNhalfflat}),
under conditions similar to the TiZrV samples above. The SEY max
is reduced from 1.5 to 1.1 and is similar to SEY obtained after
electron conditioning \cite{lepimpec:Nima2005}

The main peaks of the residual gas present in the measuring system
during ion exposure are listed in Table.\ref{tabN2pressureTINal}.
Ar seen in the spectrum could also be released during ion
bombardment of the samples. Ar is also used as a medium gas to
deposit TiN or NEG films, and is sometimes implanted. However, XPS
did not show any traces of Ar in our thin films. The pumping of
nitrogen at this level of pressure is sufficient to release noble
gas buried in the ion pump cathode plates. Hence, the Ar seen must
come solely from the pump.

\begin{table}[tbph]
\centering \caption{System partial pressures, during N$_2^+$
bombardment, P$_t \sim$7.10$^{-9}$  Torr}
\begin{tabular}{|c|c|c|c|c|c|c|c|c|}
  \hline
  Mass (amu) & 2 & 4 & 12 & 14 & 15 & 16 & 28 & 40 \\
  \hline
  Pressure (10$^{-11}$ Torr) & 200 & 3 & 3 & 10 & 1.5 & 3 & 100 & 3 \\
  \hline
\end{tabular}
\label{tabN2pressureTINal}
\end{table}

The N$_2^+$ bombardment removes water, hydrocarbons, and oxygen
from oxynitride. Because water and hydrocarbons are a high SEY
contaminant, their removal reduces the SEY \cite{Halbritter:84}.
The XPS chemistry of the TiN was monitored during bombardment and
the main result is presented below, Fig.\ref{figXPSTiN}. As
the surface contamination/oxidation is removed, the Ti~2p and N~1s
peak intensities rise. On the contrary, the C~1s and O~1s peaks
intensities decrease. The location of all peaks remains unchanged.

\begin{figure}[tbph]
\begin{center}
\includegraphics[width=0.6\textwidth,clip=]{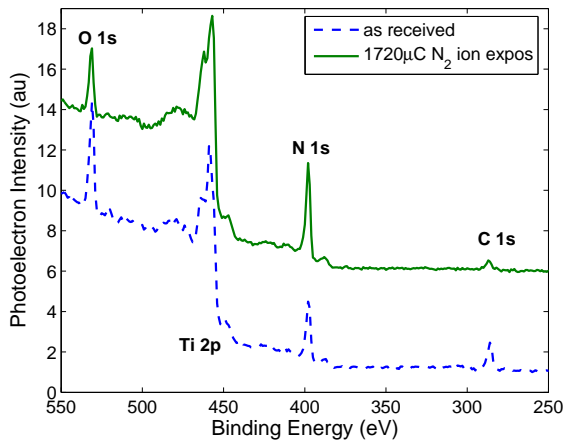}
\end{center}
\caption{XPS spectra of coated TiN/Al, as received (dashed line)
and after 1720~$\mu$C N$_2^+$ ion bombardment (solid). Spectra are
vertically displaced for clarity.}
\label{figXPSTiN}
\end{figure}

Ion bombardment by a heavier ion (Ar at 500~eV) could introduce N
vacancies \cite{Prieto:1995}, and CO$_2$ gas could play this role.
Fortunately, in a baked accelerator, even under the presence of
synchrotron radiation, its presence is negligible compared to
hydrogen and CO. No extra XPS peaks, for example due to interstitial N, were
detected in the spectrum after N$_2^+$ exposure. This does not
mean that vacancies weren't being produced, but they could have been filled by
the nitrogen coming from the ion beam.

\section{Conclusion}

TiCN SEY results, as-received  and  after heating,  do  not differ
from  results obtained on TiN \cite{lepimpec:Nima2005}. We expect
that, according to the literature, there will also be no advantage
in using a simpler TiC coating \cite{Garwin:1987}, because it has
a similar SEY and behavior upon heating. The important part of
this study compares the efficiency of 250~eV ion-bombardment with
electron-conditioning. Table.\ref{tabSEYiondose} shows that a few
micro-coulombs per mm$^2$ of ions is sufficient to bring the SEY
max below 1.3. In order to reach such a value with 130~eV
electrons, a few thousand times higher dose is needed,
Table.\ref{tabSEYelectrondose}.

\begin{table}[tbph]
\centering \caption{$\delta_{max}$ reduction due to 250~eV ion
conditioning}
\begin{tabular}{|c|c|c|c|c|}
  \hline
  Thin Film & ion species & $\delta_{max}$ & Energy$_{max}$ & Dose $\mu C/mm^2$ \\
  \hline
  TiCN & H$_2$ & 1.29 & 280 & 1.11 \\
  TiN & N$_2$ & 1.09 & 260 & 3.39 \\
  TiZrV & N$_2$ & 1.15 & 300 & 2.29 \\
  \hline
\end{tabular}
\label{tabSEYiondose}
\end{table}

\begin{table}[tbph]
\centering \caption{$\delta_{max}$ reduction due to 130~eV
electron conditioning. Data obtained at 23 deg from normal
incidence \cite{lepimpec:Nima2005}}
\bigskip
\begin{tabular}{|c|c|c|c|}
  \hline
  Thin Film & $\delta_{max}$ & Energy$_{max}$ & Dose $\mu C/mm^2$ \\
  \hline
  TiZrV/SS(CERN) & 1.21 & 300 & 11233 \\
  TiZrV/Al (SAES) & 1.07 & 370 & 8425 \\
  TiN/Al (LBL) & 1.11 & 290 & 6829 \\
   TiN/Al (BNL) & 1.01 & 380 & 6529 \\
   TiN/SS (BNL) & 1.01 & 290 & 7720 \\
  \hline
\end{tabular}
\label{tabSEYelectrondose}
\end{table}

It is also not excluded that heavier ions (N$_2^+$ vs. H$_2^+$)
may be more efficient in conditioning than lighter ions, at least
in the case of TiZrV (Fig.\ref{figTiZrVSEY}). For a dose of
0.96~$\mu C/mm^2$, the $\delta_{max}$ of the TiZrV is already
below 1.2. It might also be interesting to now study the effect of
CO ion-bombardment vs. CO residual gas adsorption. Based on
physical sputtering, conditioning should occur. However, the
possibility of surface oxidation by dissociation of the CO could
increase the SEY. Exposure to background CO alone does increase
the SEY \cite{Scheuerlein:2002}; however, there will be
competition for the case of H$_2^+$/CO$^+$ ion-sputtering vs. CO
vacuum recontamination, as is the case for a circulating positron
beam.

\section{Acknowledgments}

This work would not have been carried out without the strong
support of Prof T.~Raubenheimer and the full ILC team at SLAC. The
authors would also like to thanks Prof A.~Wrulich, at PSI, for his
interest in the R\&D of the future linear accelerator.  This work
is supported by the U.S. Department of Energy under contract
number DE-AC02-76SF00515.

%
%


\begin{thebibliography}{10}

\bibitem{lepimpec:Nima2005}
{F. Le Pimpec, F. King, R.E. Kirby, M. Pivi}.
\newblock {Properties of TiN and TiZrV Thin Film as a Remedy Against Electron
  Cloud}.
\newblock {\em Nuclear Instruments and Methods in Physics Research A}, 551
  (2-3):187--199, 2005.

\bibitem{lepimpec:jvsta2005}
{F. Le Pimpec, F. King, R.E. Kirby, M. Pivi}.
\newblock {Electron Conditioning of Technical Aluminium Surfaces: Effect on the
  Secondary Electron Yield}.
\newblock {\em Journal of Vacuum Science and Technology}, A (23):1610, 2005.

\bibitem{lanfawang}
L. Wang, SLAC private communication.

\bibitem{Hopf:2003}
{C. Hopf, A. von Keudell and W. Jacob}.
\newblock {Chemical sputtering of hydrocarbon films}.
\newblock {\em J. Appl. Phys.}, 94:2373, 2003.

\bibitem{Halbritter:84}
{J. Halbritter}.
\newblock {On Changes of Secondary Emission by Resonnant Tunneling via
  Adsorbates}.
\newblock {\em Journal de Physique}, 45:C2--315, 1984.

\bibitem{Hoyt:Pac95}
{E. Hoyt, M. Hoyt, R.E. Kirby, C.~Perkins, D.~Wright and
A.~Farvid}.
\newblock {Processing of OFE copper beam chambers for PEP-II high energy ring}.
\newblock In {\em {PAC95, IEEE Proceedings v.3, 2075}}, 1995.

\bibitem{kirby:1980}
{R.E.~Kirby, C.S.~McKee and L.V.~Renny}.
\newblock {Faceting of Cu(210) and Ni(210) by activated nitrogen}.
\newblock {\em Surface Sci.}, 97:457, 1980.

\bibitem{crc}
David~R. Lide, editor.
\newblock {\em {Handbook of Chemistry and Physics}}.
\newblock 74$^{th}$ edition. CRC~PRESS, 1994.

\bibitem{Garwin:1987}
{E.L Garwin F.K.~King R.E.~Kirby and O. Aita}.
\newblock {Surface Properties of Metal-Nitride and Metal-Carbide films
  deposited on Nb for radio-frequency superconductivity}.
\newblock {\em Journal of Applied Physics}, 61(3), 1987.

\bibitem{he:EPAC04}
{P. He et al.}
\newblock { Secondary Electron Emission Measurements for TIN coatings on the
  Stainless Steel of SNS Accumulator Ring Vacuum Chamber}.
\newblock In {\em {EPAC 2004}}, 2004.
\newblock {SLAC-PUB-10570}.

\bibitem{Hilleret:HPC02}
{N. Hilleret}, 2002.
\newblock
  {http://laser.jlab.org/devlore/filebin/6701/HPC2002/talks/ hilleret.pdf}.

\bibitem{calder:1986}
{R. Calder, G. Dominichini, N.~Hilleret}.
\newblock {Influence of various vacuum surface treatments on the secondary
  electron yield of Niobium}.
\newblock {\em Nuclear Instruments and Methods in Physics Research B}, B13:631,
  1986.

\bibitem{Padamsee:1979}
{H. Padamsee and A. Joshi}.
\newblock {Secondary electron emission measurements on materials used for
  superconducting microwave cavities}.
\newblock {\em {Journal of Applied Physics}}, 50(2):1112, 1979.

\bibitem{Scheuerlein:2002}
C.~Scheuerlein.
\newblock {The Activation of Non-evaporable Getters Monitored by AES, XPS,
  SSIMS and Secondary Electron Yield Measurements}.
\newblock Technical report, 2002.
\newblock {CERN- THESIS- 2002- 026}.

\bibitem{Chiggiato}
P. Chiggiato, CERN private communication.

\bibitem{Yulin}
Y. Li, Cornell University private communication.

\bibitem{lozano:2000}
{M. P. Lozano and J. Fraxedas}.
\newblock {XPS Analysis of the Activation Process in Non-Evaporable Getter Thin
  Films}.
\newblock {\em {Surface and Interface Analysis}}, 30:623, 2000.

\bibitem{Welch:1472}
{K.M. Welch}.
\newblock {Low Pressure Crossed Field Vacuum Sputtering of Thin Films for
  Multipactor Suppression Using a Simple Diode Array}.
\newblock Technical report, SLAC-Pub-1472, 1974.

\bibitem{kirby:2001}
{R.E.~Kirby, F.K.~King}.
\newblock {Secondary Emission Yield from PEP-II accelerator material}.
\newblock {\em Nuclear Instruments and Methods in Physics Research A}, A469,
  2001.

\bibitem{galan:esa03}
{L. Gal\'an, et al.}
\newblock {Surface Treatment and Coating for the Reduction of Multipactor and
  Passive Intermodulation (PIM) Effects in RF Components}.
\newblock In {\em {4$^{th}$ International Workshop on Multipactor, Corona and
  PIM in Space Hardware}}, 2003.

\bibitem{Prieto:1995}
{P.~ Prieto and R.E.~Kirby}.
\newblock {X-ray photoelectron spectroscopy study of the diference between
  reactively evaporated and direct sputter-deposited TiN films and their
  oxidation prperties}.
\newblock {\em Journal of Vacuum Science and Technology}, A13(6), 1995.

\end{thebibliography}

\clearpage
\listoftables
\newpage
\listoffigures

\end{document}